\input{aipcheck}

\documentclass[
    ,final            
  ]
  {aipproc}

\layoutstyle{6x9}

\begin{document}
\input{psfig.sty}
\title{Update of  CDF Results on Diffraction}
\author{Konstantin Goulianos}
{address=
{The Rockefeller University, 
1230 York Avenue, New York, NY 10021, U.S.A.\\
{\rm (Presented on behalf of the CDF Collaboration)}}
}



\begin{abstract}
The diffractive program of the CDF Collaboration 
at the Fermilab Tevatron $\bar pp$ Collider  
is reviewed with emphasis on recent results from Run-II and future prospects. 
\end{abstract}

\maketitle

\vspace{-17em}

\begin{center}Presented at DIS-2005, XIII$^{th}$ International Workshop on Deep Inelastic Scallering,\\
April 27 - May 1 2005, Madison, WI, U.S.A.
\end{center}
\vspace{13em}

Diffractive $\bar pp$ interactions are characterized by the 
presence of at least one large rapidity gap, defined as a  region 
of pseudorapidity~\cite{rapidity} devoid of particles.
A diffractive rapidity gap, which  may 
be forward (adjacent to a leading nucleon) 
or central, is presumed to be formed by the exchange 
of a {\em Pomeron}~\cite{Regge}, which in QCD is a  color singlet 
quark/gluon object with vacuum quantum numbers. 
Diffraction in which there is a high momentum-transfer 
partonic scattering in the event in addition to the rapidity gap 
is referred to as {\em hard diffraction}. 
In this paper, we briefly review the results on diffraction 
obtained by the Collider Detector at Fermilab (CDF) in Run-I 
(1992-1995), present an update of results from Run-II, 
which is in progress, and discuss future prospects.
\section{Run-I RESULTS}
In addition to measuring $\bar pp$ elastic, single diffraction (SD), 
and total cross sections at $\sqrt s=540$ and 1800 GeV, CDF studied
several soft and hard diffraction processes at 1800 GeV, 
and in some cases at $\sqrt s =630$ GeV~\cite{lathuile}. 
Soft processes studied include:
\begin{center}
\begin{tabular}{lll}
{\bf DD}&Double Diffraction&$\bar{p}p\rightarrow X+{\rm gap}+Y$\\
{\bf DPE}&Double Pomeron Exchange&$\bar{p}p\rightarrow \bar{p}+{\rm gap}+X+{\rm gap}+p$\\
{\bf SDD}&Single $\oplus$ Double Diffraction&
$\bar{p}p\rightarrow \bar{p}+{\rm gap}+X+{\rm gap}+Y$\\
\end{tabular}
\end{center}
In hard diffraction CDF measured SD dijet, $W$, $b$-quark and
$J/\psi$, DD dijet, 
and DPE dijet production.
Schematic diagrams and event topologies for representative  
processes are shown in Fig.~\ref{fig:diagrams}.
 
\vspace{-1em}
\begin{minipage}[t]{0.5\textwidth}
\phantom{xxx}
\centerline{\bf\sc SOFT DIFFRACTION}
\vspace{1em}
\hspace{-0.5em}\includegraphics[width=0.95\textwidth]{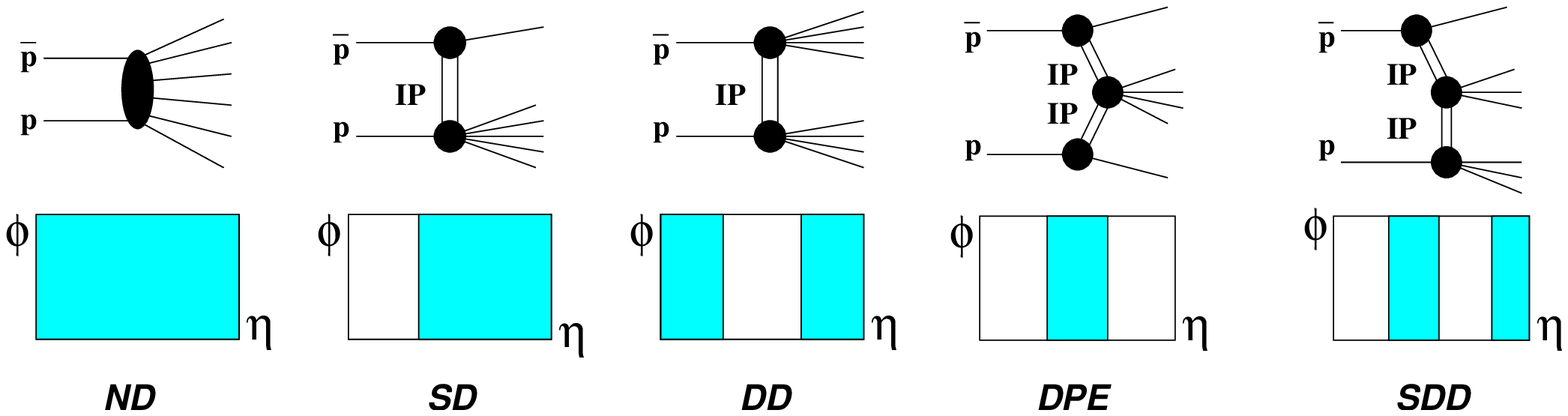}
\end{minipage}
\hspace{0.02\textwidth}
\begin{minipage}[t]{0.5\textwidth}
\phantom{xxx}
\centerline{\bf\it\sc HARD DIFFRACTION}
\vspace{1em}\vspace{-0.5ex}
\hspace{-1em}\includegraphics[width=1.1\textwidth]{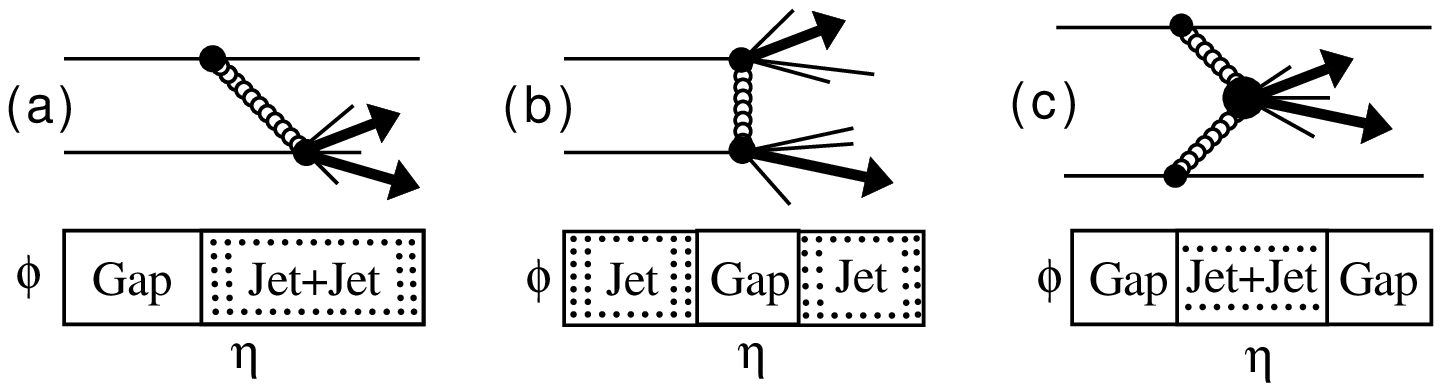}
\vspace{-22em}
\end{minipage}

\begin{figure}[h]
\caption{Schematic diagrams and $\eta$-$\phi$ topologies of representative 
diffractive processes studied by CDF. The shaded areas represent regions of 
pseudorapidity in which there is particle production.}
\label{fig:diagrams}
\end{figure}

Two types of hard diffraction results were obtained in Run-I: 
diffractive to non-diffractive cross 
section ratios using the rapidity gap signature 
to select diffractive events 
and diffractive to non-diffractive structure function ratios  using a 
Roman Pot Spectrometer (RPS) to trigger on leading antiprotons. 
The results exhibit regularities in normalization
and factorization properties that point to
the QCD character of diffraction (see~\cite{lathuile}).

At $\sqrt s=$1800 GeV, the SD/ND ratios (gap fractions)   
for dijet, $W$, $b$-quark, and $J/\psi$ production, as well the ratio of
DD/ND dijet production, are all $\approx 1\%$.
These ratios are suppressed relative to standard QCD inspired theoretical 
expectations ({\em e.g.} 2-gluon exchange) 
by a factor of $\sim$10, which is comparable to that
observed in soft diffraction relative to Regge theory expectations.
This suppression represents a severe breakdown of QCD 
factorization. It is, however, interesting to note that except for the 
overall suppression in normalization factorization approximately holds 
at fixed $\sqrt s$.

Another interesting aspect of the Run-I results is that 
ratios of two-gap to one-gap cross sections 
for both soft and hard processes appear to obey factorization. This feature 
of the data provides both a clue to understanding diffraction and 
a tool for diffractive studies using processes with 
multiple rapidity gaps~\cite{KG:here}.

\section{Run-II Program}
The goal of the Run-II diffractive program of CDF is twofold:
(a) to obtain results that could help decipher the QCD 
nature of the Pomeron, such as dependence of the diffractive structure 
function (DSF) on $Q^2$, $x_{Bj}$, $t$, 
and $\xi$ (fractional momentum loss of the diffracted nucleon), 
and (b) to measure exclusive production rates 
(dijet, $\chi_c^0$, $\gamma\gamma$), 
which could to be used to establish 
benchmark calibrations for 
exclusive Higgs production at LHC~\cite{KMR}.  
Preliminary results from data collected at $\sqrt s=1.96$ GeV  
confirm the Run-I DSF results~\cite{lathuile,MGmoriond}.
New in Run-II are the measurement of the 
$Q^2$ dependence of the DSF obtained 
from dijet production and 
limits on exclusive production rates.  

\vglue -0.5em
\hspace*{-0.8em}\begin{minipage}[t]{0.5\textwidth}
\phantom{xxx}
\hspace{-1em}\includegraphics[width=1.05\textwidth]{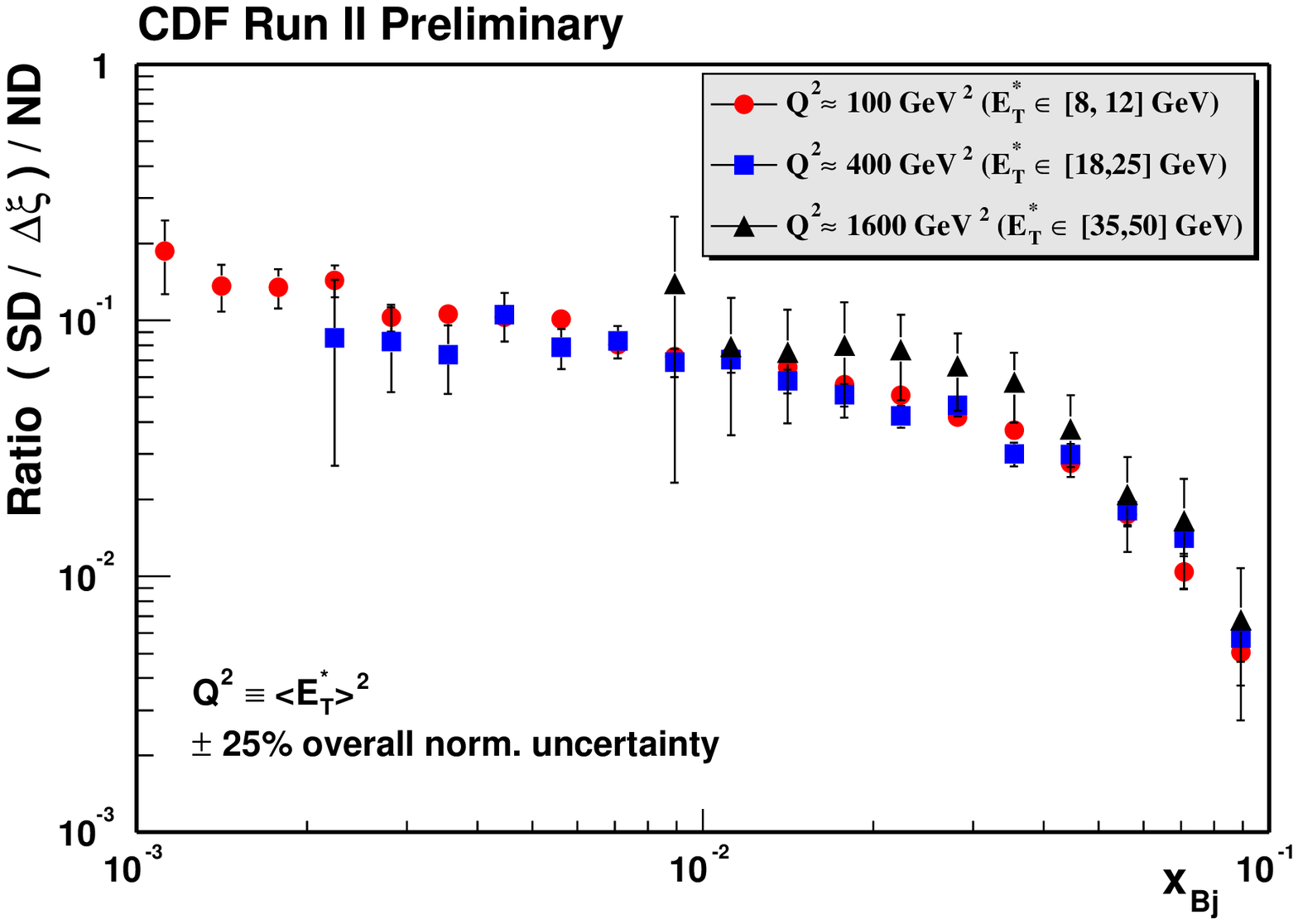}
\end{minipage}
\hspace{0.0ex}
\begin{minipage}[t]{0.5\textwidth}
\phantom{xxx}
\hspace{-1em}\includegraphics[width=1.14\textwidth]{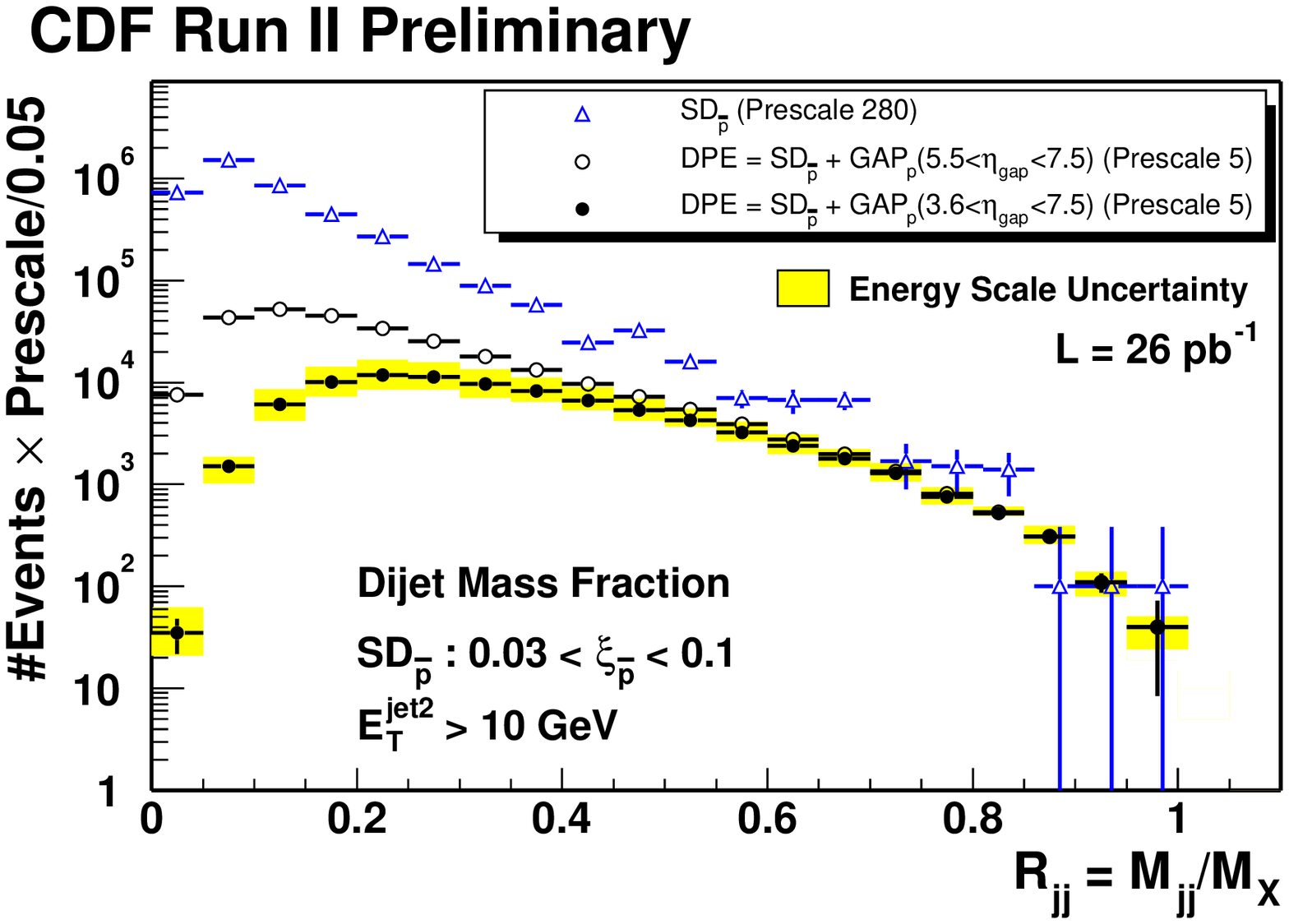}
\end{minipage}

\vglue -2ex
\begin{figure}[ht]
\caption{({\em left}) Ratio of SD/$\Delta\xi_{\bar{p}}$
over ND rates  obtained
from dijet data at various $Q^2$ ranges; {(\em right})  
ratio of dijet mass to 
total mass ``visible'' in the calorimeters for dijet production in events 
with a leading antiproton within $0.3<\xi_{\bar{p}}<0.1$ and various 
gap requirements on the proton side: ({triangles}) no gap requirement, 
({open circles}) gap in 
$5.5<\eta<7.5$, and ({filled circles}) gap in region $3.5<\eta<7.5$.}
\label{fig:Q2Mjj}
\end{figure}
 
\subsection{The diffractive structure function}
In Fig.~\ref{fig:Q2Mjj}~{(\em left)},
the ratio of SD/ND rates, which in LO QCD and at fixed $x_{Bj}$ 
is equal to the ratio of the corresponding structure functions, 
shows no appreciable $Q^2$ dependence. 
This result was foreseen in the renormalization model~\cite{newapproach}. 
in which the diffractive structure function is basically the low-$x$ 
($x<\xi$) structure function of the diffracted nucleon. 
More data are currently being analyzed to improve the statistics 
of this measurement.

Data are at hand and analyses are in progress for the measurement of 
the $t$, $\xi$, and flavor dependence of the DSF 
using dijet, $W$, and $J/\psi$ production. In addition, factorization 
will be tested more accurately than in Run-I by comparing the 
DSFs obtained from dijet production in SD and DPE.    

\subsection{Exclusive production}
\subsubsection{Exclusive dijet production}
The search for exclusive dijet production is based on measuring the 
dijet mass fraction $M_{jj}$, 
defined as the mass of the two leading jets in an event 
divided by the total mass 
reconstructed from all the energy observed in all calorimeters.
Fig.~\ref{fig:Q2Mjj} ({\em right}) shows $M_{jj}$ distributions for events 
with different selection criteria. The signal from exclusive dijets
is expected to be concentrated in the region of $R_{jj}>0.8$,
with values of $R_{jj}<1$ being caused by measurement resolution effects and 
final state radiation. Of course, background events from 
inclusive DPE production, 
$\bar p p\rightarrow (\bar p+gap)+JJ+X+gap$, are expected to contribute to the 
entire $M_{jj}$ region. 

\vglue -0.5em
\hspace*{-1em}\begin{minipage}[t]{0.5\textwidth}
\phantom{xxx}
\includegraphics[width=1.1\textwidth]{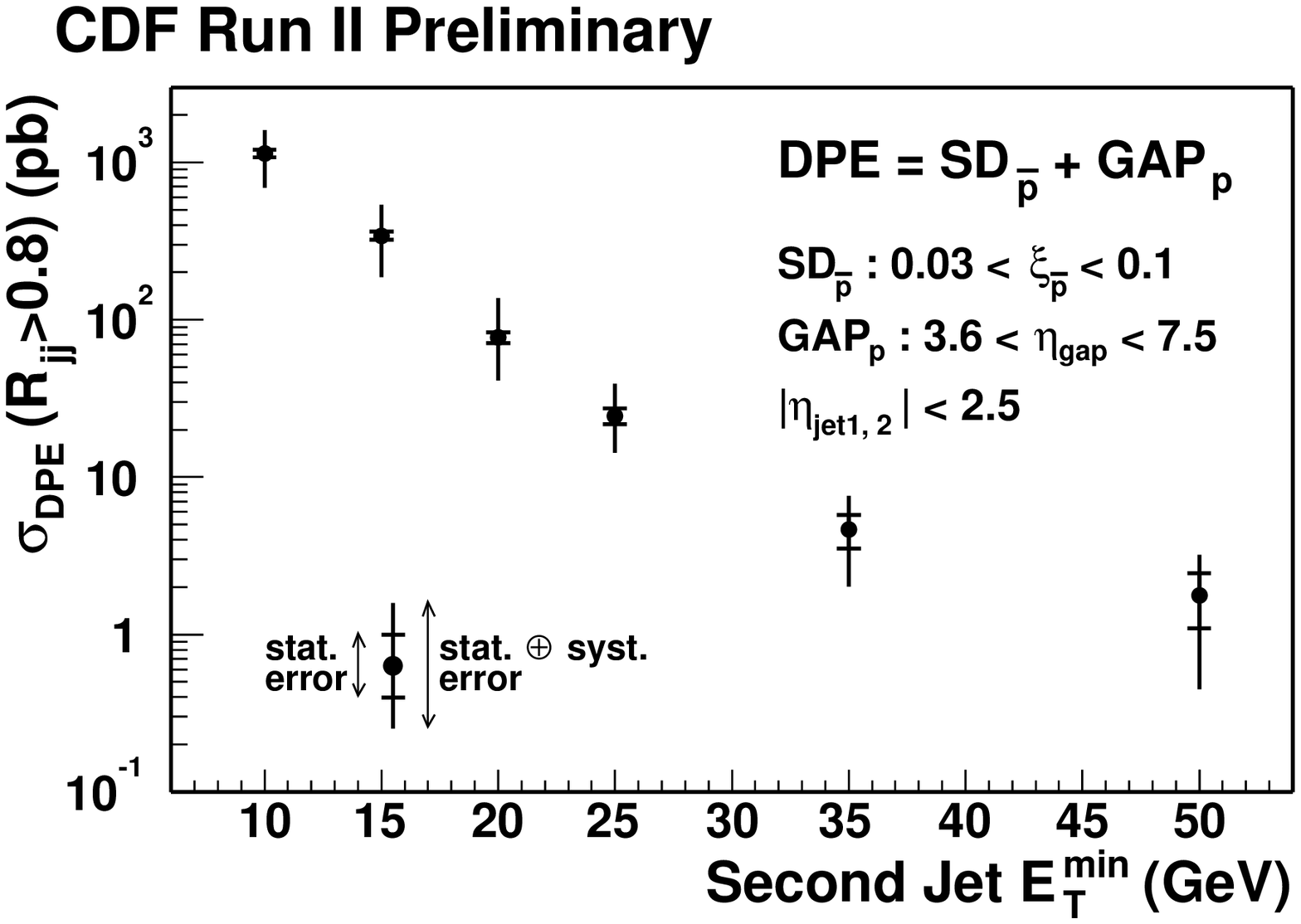}
\end{minipage}
\hspace{2ex}
\begin{minipage}[t]{0.5\textwidth}
\phantom{xxx}
\vspace{-0.3em}
\includegraphics[width=1\textwidth]{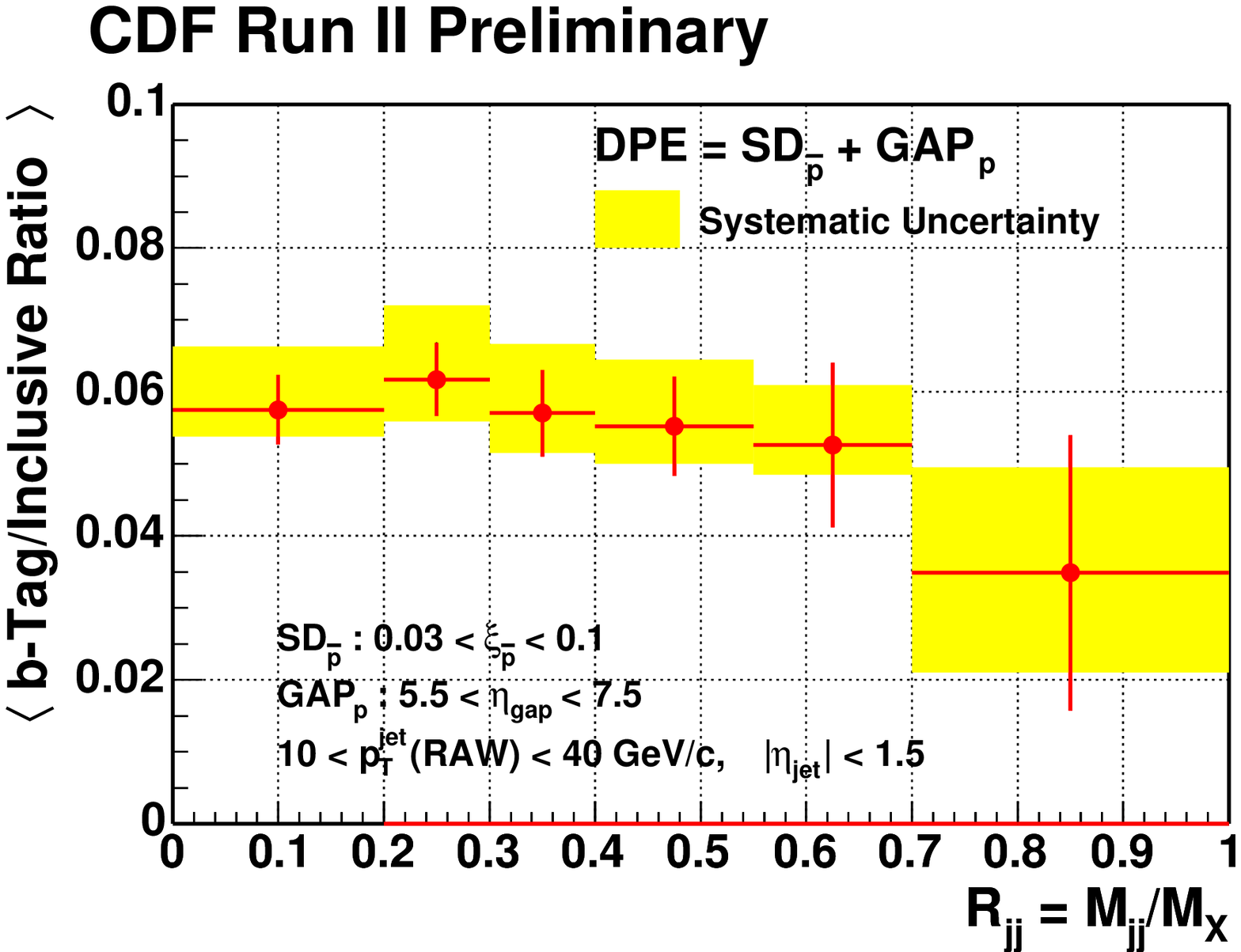}
\end{minipage}

\vglue -1em
\begin{figure}[ht]
\caption{
({\em left}) 
Dijet production cross sections for $R_{jj}>0.8$ in DPE events 
as a function of $E_T^{min}$, the $E_T$ of the next to the highest 
$E_T$ jet; 
({\em right}) the ratio of $b$-tagged to all jets in the DPE 
dijet event sample versus the dijet mass fraction.   
}
\label{fig:exclusiveJJ}
\end{figure}

Since no peak is observed at $R_{jj}>0.8$ in 
Fig.~\ref{fig:Q2Mjj} ({\em right)}, CDF reports production 
cross sections for events with $R_{jj}>0.8$, which could be used 
as upper limits for exclusive production. 
Figure~\ref{fig:exclusiveJJ}~({\em left)} 
shows such cross sections for various kinematic cuts plotted versus 
$E_T^{min}$, the next to leading jet $E_T$. These cross 
sections agree, within errors, with recent predictions for exclusive dijet 
production~\cite{KMR}. Thus, for these predictions to be correct, 
the background would have to vanish as $R_{jj}\rightarrow 1$. While this is 
guaranteed by the $J_z=0$ selection rule for leading order 
$gg\rightarrow q\bar q$ jets of
$m_q<<M_{jet}$, Monte Carlo (MC) simulations are used to deal 
with the dominant $gg\rightarrow gg$ process. 
To avoid using simulations $q\bar q$ events could be used to estimate
the background and this could be done using dijet events in which at least 
one of the jets is $b$-tagged. 
Figure~\ref{fig:exclusiveJJ}~({\em right}) shows the ratio of 
$b$-tagged to inclusive dijet events versus dijet mass fraction.
A suppression is observed as $M_{jj}\rightarrow 1$, as 
would be expected if there were exclusive dijets in the sample. 
However, background still may exist 
from the gluon splitting process 
$gg\rightarrow g+g(\rightarrow b\bar b)$. This background could be 
practically eliminated if both jets were required to be $b$-tagged. 
Presently, more data are being 
collected with an unprescaled $b$-tagged dijet trigger 
to yield a large sample of double-$b$-tagged dijet events to 
measure the rate for exclusive production in a low background 
environment. 

\subsubsection{Exclusive $\chi_c^0$ production}
CDF has reported an upper limit of 
$49\pm 18\mbox{ (stat)}\pm 39\mbox{ (syst)}$~pb
for exclusive $\chi_c^0$ production
from a search for $J/\psi+\gamma$ events from 
$\bar pp\rightarrow \bar p+\chi_c^0(\rightarrow J/\psi+\gamma\rightarrow \mu\mu+\gamma)+\bar p$. 
Theoretical predictions of $\sim 70$~pb  
have recently been revised to $\sim 50$~pb~\cite{KMR}. More data, 
collected with a dedicated trigger, are currently being analyzed. 
 
\section{Conclusions}
A comprehensive program of measurements of the diffractive 
structure function and of exclusive diffractive production 
is currently under way at CDF aiming at deciphering the 
QCD nature of diffraction and at providing benchmark calibrations for 
estimating rates for diffractive Higgs production at the LHC.


\begin{thebibliography}{99}
\bibitem{rapidity}We use rapidity, $y=\frac{1}{2}\frac{E+p_L}{E-p_L}$, and 
pseudorapidity, $\eta=-\ln\tan\frac{\theta}{2}$, interchangeably, 
since in the kinematic region of interest they are approximately equal.
\end{thebibliography}
\end{document}